\documentclass[10pt,final,conference]{IEEEtran}
\IEEEoverridecommandlockouts

\PassOptionsToPackage{hyphens}{url}
\usepackage[table,xcdraw]{xcolor}
\usepackage{hyperref, url}
\usepackage{amsmath,amssymb,amsfonts}
\usepackage{algorithm}
\usepackage{graphicx}
\usepackage{algpseudocode}
\algnewcommand{\LeftComment}[2][\algorithmicindent]{\Statex \hspace{#1} \(\triangleright\) #2}

\usepackage{multirow}

\usepackage{textcomp}
\usepackage{xcolor}
\usepackage[T1]{fontenc}
\def\BibTeX{{\rm B\kern-.05em{\sc i\kern-.025em b}\kern-.08em
    T\kern-.1667em\lower.7ex\hbox{E}\kern-.125emX}}

\begin{document}

\title{GitHub Copilot: the perfect Code compLeeter?}

\author{\IEEEauthorblockN{1\textsuperscript{st} Ilja Siroš}
\IEEEauthorblockA{\textit{COSIC, KU Leuven} \\
Leuven, Belgium \\
ilja.siros@esat.kuleuven.be}
\and
\IEEEauthorblockN{2\textsuperscript{nd} Dave Singelée}
\IEEEauthorblockA{\textit{COSIC, KU Leuven} \\
Leuven, Belgium \\
dave.singelee@esat.kuleuven.be}
\and
\IEEEauthorblockN{3\textsuperscript{rd} Bart Preneel}
\IEEEauthorblockA{\textit{COSIC, KU Leuven} \\
Leuven, Belgium \\
bart.preneel@esat.kuleuven.be}
}

\maketitle



\begin{abstract}
This paper aims to evaluate GitHub Copilot's generated code quality based on the LeetCode problem set using a custom automated framework. We evaluate the results of Copilot for 4 programming languages: Java, C++, Python3 and Rust. We aim to evaluate Copilot's reliability in the code generation stage, the correctness of the generated code and its dependency on the programming language, problem's difficulty level and problem's topic. In addition to that, we evaluate code's time and memory efficiency and compare it to the average human results. In total, we generate solutions for 1760 problems for each programming language and evaluate all the Copilot's suggestions for each problem, resulting in over 50000 submissions to LeetCode spread over a 2-month period. We found that Copilot successfully solved most of the problems. However, Copilot was rather more successful in generating code in Java and C++ than in Python3 and Rust. Moreover, in case of Python3 Copilot proved to be rather unreliable in the code generation phase. We also discovered that Copilot's top-ranked suggestions are not always the best. In addition, we analysed how the topic of the problem impacts the correctness rate. Finally, based on statistics information from LeetCode, we can conclude that Copilot generates more efficient code than an average human.
\end{abstract}


\section{Introduction}
In June $2021$, GitHub released Copilot \cite{Copilot}, an ``AI pair programmer" that can generate the code for different languages given some context, such as comments, function names or surrounding code. In any given situation, Copilot will typically suggest multiple ways to complete the code. Those solutions are ranked from top to bottom, from most to least likely to be correct. However, code generated by GitHub Copilot might not always work or even make sense. According to GitHub, about $26\%$ of all Copilot's suggestions have been accepted by the users. This paper aims to systematically evaluate the correctness and efficiency of the code generated by Copilot, using a fully automated evaluation process.

For this purpose, we use the LeetCode \cite{LeetCode} problem set, which includes many programming problems on various topics and is typically used by software developers to improve their programming skills or to prepare for job interviews. Each problem on LeetCode is assigned a difficulty level (easy, medium and hard), based on the median time required to solve the problem: problems taking less than 15 minutes are classified as easy, medium problems are those taking less than 40 minutes, and hard problems take more than 40 minutes. However, this assignment is subjective and according to one of the answers posted on the LeetCode forum~\cite{LeetCode_questions_difficulty}, it is chosen by the person who created the problem.

\begin{figure}
    \centering
    \includegraphics[scale=0.25]{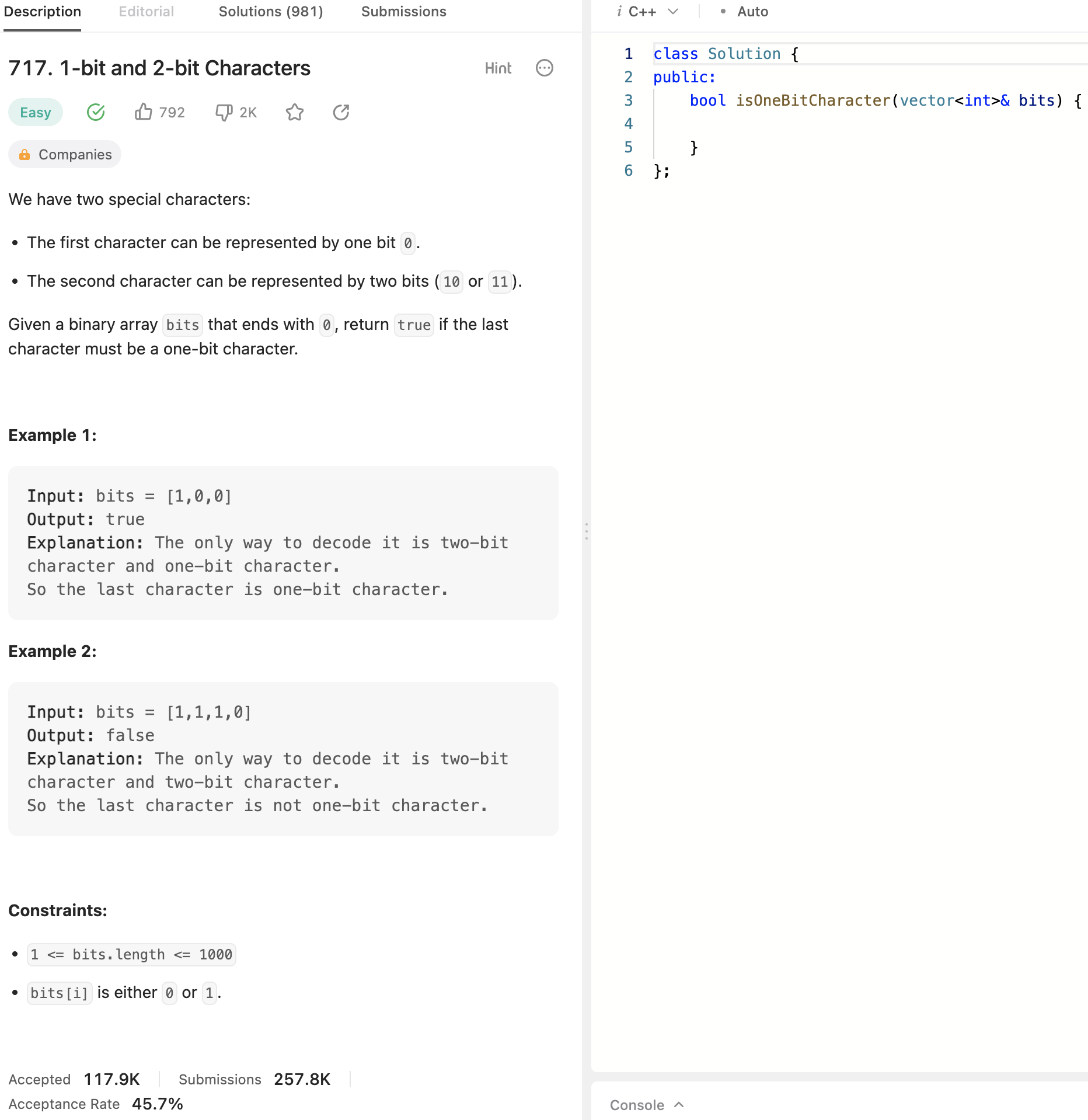}
    \caption{Example of a LeetCode problem. On the left part of the figure, the problem description is shown and some examples are mentioned together with the constraints imposed on the solution by LeetCode. The right part of the figure shows the solution template for the chosen programming language, containing the function that needs to be implemented.}
    \label{fig:LeetCode_problem}
\end{figure}

An example of a LeetCode problem is shown in Figure \ref{fig:LeetCode_problem}. Every LeetCode question comes with tests for different programming languages. If the solution to the problem is correct, LeetCode provides information about its efficiency compared to all other correct solutions that it has already received. Efficiency is measured in terms of runtime and memory consumption. We use both of these metrics to evaluate Copilot-generated code.

Overall, we used Copilot to generate solutions to all LeetCode problems that are publicly available (i.e. do not require a premium account), support at least one of programming languages chosen by us, and do not require the implementation of more than one function. The latter limitation simplifies the automation process without sacrificing too many problems, as most LeetCode questions require the implementation of a single function. In total, we ended up with 1760 problems per programming language. For each problem, we asked Copilot to generate as many suggestions as possible and tested all of them for correctness, time and memory efficiency. It allows us to investigate how the ranking of Copilot's suggestion correlates with correctness as well as to compare the code generated by Copilot compares to all other submissions from LeetCode.

In our work, we explicitly address the following research questions:
\begin{itemize}
    \item{\textbf{RQ1}} How reliable is Copilot in the code generation phase? More specifically, how often does Copilot fail to generate any code for a given problem and what is the average number of generated suggestions depending on the programming language?
    \item{\textbf{RQ2}} How correct are Copilot’s suggestions? More specifically, how does the correctness of Copilot’s suggestions depend on the programming language, the difficulty of the problem, the rank of the suggestion in the code generation phase and the topic of the problem?
    \item{\textbf{RQ3}} How does the time and memory efficiency of the correctly generated suggestion compare to that of the average human submission?
\end{itemize}

\section{Background and Related Work} \label{sec:RelatedWork}
Since Copilot's appearance in $2021$, there has been a significant amount of research work done exploring the correctness, security and understandability of Copilot's generated code.

Pearce et al.~\cite{pearce2021asleep} evaluated the security of Copilot-generated code, assessing the presence of the top $25$ most dangerous software weaknesses in the generated code using the list of Common Weaknesses Enumeration~\cite{CWE}. They showed that developers should be alert when using Copilot as sometimes it produces a vulnerable code and stated that ideally, Copilot should be used along with a security-aware tooling to minimize the risk of introducing security vulnerabilities.

Imai~\cite{10.1145/3510454.3522684} conducted an empirical experiment to compare AI with human participants in a natural software development environment, measuring productivity and code quality with and without Copilot. Arghavan et al.~\cite{dakhel2023github} also conducted an empirical study on the performance and functionality of Copilot's suggestions for fundamental algorithmic problems and compared Copilot's solutions with human solutions on a dataset of Python programming problems. 
In these two papers, the authors concluded that even though Copilot helps to increase developers' productivity and can generate optimal solutions for some fundamental problems, the generated code's quality highly depends on the conciseness and depth of the given prompt.


Yetistiren et al.~\cite{10.1145/3558489.3559072} and Nguyen et al.~\cite{Copilot_Eval} evaluated the correctness of Copilot solutions for $146$ problems (Python only) in HumanEval dataset~\cite{humaneval} and $33$ LeetCode problems for $4$ programming languages (Python3, Java, JavaScript, C), respectively. However, both works lacked full automation in the evaluation process, leading to a small dataset and few generated solutions. Hence, their studies cannot be considered comprehensive enough to make definite conclusions about the correctness of Copilot-generated code.

Döderlein et al.~\cite{döderlein2023piloting} expanded the research by evaluating the correctness of generated solutions by Codex and Copilot for $146$ problems from the HumanEval dataset as well as $300$ LeetCode problems ($100$ for each difficulty level) for $6$ programming languages (Python3, Java, JavaScript, C, C++, C\#).
However, they have only evaluated Copilot's top-ranked suggestions for the LeetCode problems, and their main research focus was on tuning the prompt and temperature to get better results from the AI models. For the Codex language model and HumanEval dataset, the authors found out that generating many suggestions drastically improves the correctness rate for the given problem, i.e. by generating more suggestions, it is more likely that one of them will be correct. If only the first suggestion was taken, the correctness rate was 22.44\%. But if the first $10$ suggestions were taken, the correctness rate was already 71.7\%. It has to be noted that they did not conduct the same experiments neither for Copilot nor for LeetCode problems.

In our research, we go beyond existing work and focus on evaluating the reliability, correctness and efficiency of Copilot-generated code using a much larger dataset containing most of the LeetCode problems. We evaluate the suggestions for $1760$ problems of different difficulty levels for each of the $4$ chosen programming languages, namely C++, Java, Python3 and Rust. Moreover, we analyse all of the Copilot's suggestions, not limiting ourselves to the top-ranked ones, which led to over $50000$ submissions to LeetCode spread over a $2$-month period. In addition we provide an information about Copilot's correctness rate depending on the topics of the LeetCode problems.

Section~\ref{subsec:comparison_with_prev_works} compares our results to those from the related works.

\section{Design}
In this section, we describe our study setup. The study was conducted using an Apple MacBook Pro A2141. Copilot version 1.79.10634 was invoked from Visual Studio Code~\cite{VS_CODE}. The script responsible for the communication with LeetCode is written in Python and relies on the python-leetcode library \cite{python-LeetCode}. To automate inter-application communication on an Apple device, AppleScript scripting language~\cite{AppleScript} was used.
Our code is open-source, and it is available at \url{https://github.com/IljaSir/CopilotSolverForLeetCode}.

\subsection{Evaluation flow}
First, we describe the automated evaluation flow, summarized in Algorithm~\ref{alg:process_flow}.
\begin{algorithm}[H]
    \caption{Process flow} \label{alg:process_flow}
    \begin{algorithmic}
        \LeftComment[0\dimexpr\algorithmicindent]{Fetch problem names from LeetCode}
        \State $ProblemNames \leftarrow $ \Call {FetchAllProblemNames}{$ $}
        \For {$ProblemName \in ProblemNames$}
            \LeftComment[1\dimexpr\algorithmicindent]{Get problem description using LeetCode API}
            \State $content \leftarrow $ \Call {GetProblemContent}{$ProblemName$}
            \LeftComment[1\dimexpr\algorithmicindent]{Save problem content to the corresponding file}
            \State $filename \leftarrow $ \Call {SaveToFile}{$ProblemName$, $content$}
            \LeftComment[1\dimexpr\algorithmicindent]{Invoke Copilot and save solutions}
            \State $CopilotSolutions \leftarrow$ \Call {RunAppleScript}{$filename$}
            
            \For {$CopilotSolution \in CopilotSolutions$}
                \LeftComment[2\dimexpr\algorithmicindent]{Submit solution to LeetCode}
                \State $SubmissionID \leftarrow$ Submit($CopilotSolution$)
                \LeftComment[2\dimexpr\algorithmicindent]{Get sumbission result and save it}
                \State $result \leftarrow$ GetResult($SubmissionID$)
                \State SaveResult($result$)
            \EndFor
            
        \EndFor
        
    \end{algorithmic}
\end{algorithm}
Here, we give more details about the evaluation flow. Overall, it can be divided into the following steps.

\subsubsection{Get the list of all LeetCode problems}
LeetCode provides an API~\cite{LeetCode_API_ALL} to extract all problem names. The problems are stored in a JSON file containing information about the problem title, difficulty and supported programming languages.

\subsubsection{Request and parse problem content}\label{subsec:request_problem_content}
\begin{figure}
    \centering
    \includegraphics[scale=0.31]{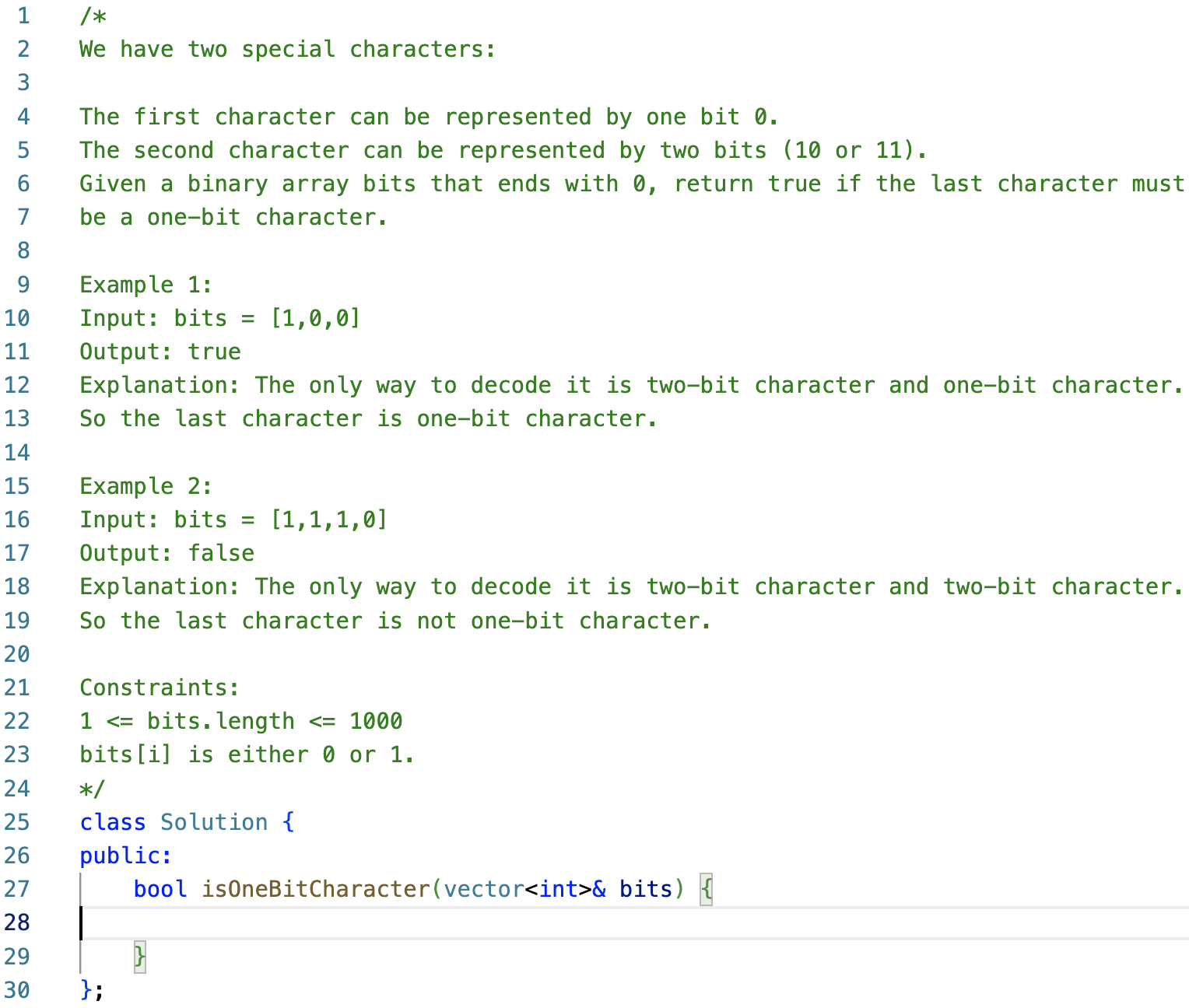}
    \caption{The code of the problem before invoking Copilot. The problem's description is added as a comment above the code template to give Copilot a context to generate a solution. Before invoking Copilot, the mouse cursor is placed inside the function that we want Copilot to generate. In this example, it is line 28.}
    \label{fig:preprocess_example}
\end{figure}

Knowing the problem name, fetching its content using LeetCode GraphQL API is straightforward.
The content includes a description of the problem and a solution template, typically just the name of the function that needs to be implemented. Note that the problem description on LeetCode sometimes includes an image with an example. However, at the time of our research Copilot could not absorb image content. Therefore, all the images were omitted, and only the text description was used.
The problem description is sanitized to remove HTML notation. After that, it is inserted above the solution template as a multi-line comment.
An example of the resulting file for the LeetCode problem from Figure \ref{fig:LeetCode_problem} is shown in Figure \ref{fig:preprocess_example}.
The resulting file is saved and opened in Visual Studio Code. 

\subsubsection{Invoke Copilot and save generated solutions} \label{subsec:invoke_copilot_and_save_solutions}
\begin{figure}
    \centering
    \includegraphics[scale=0.6]{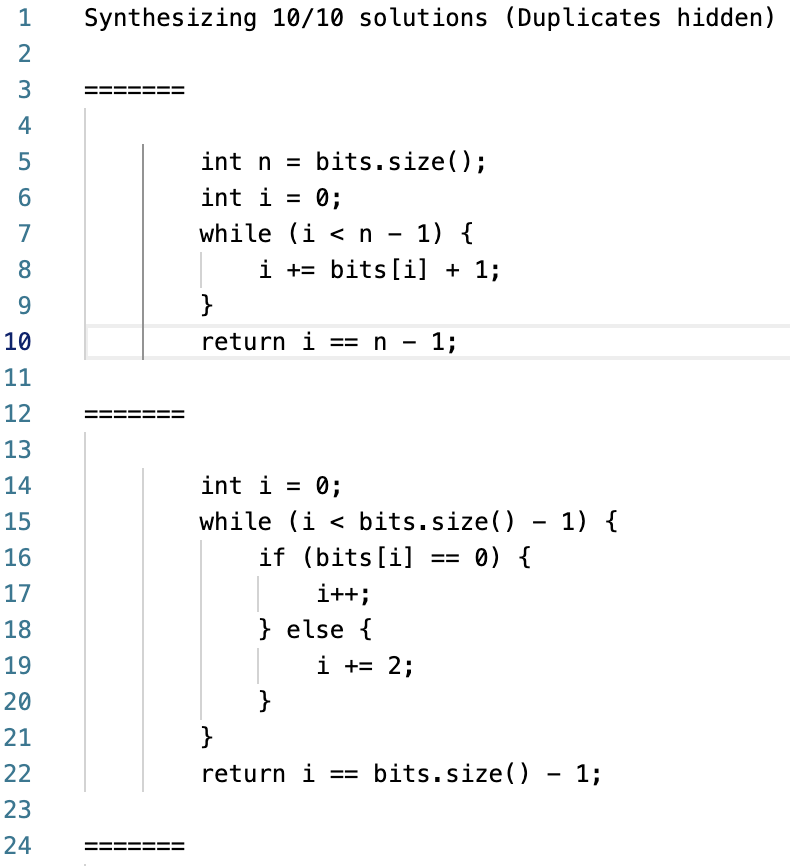}
    \caption{File with Copilot suggestions. Copilot provides multiple suggestions for the request to generate code. In this case, the number of generated suggestions is stated on line 1 and equals 10. These suggestions are ranked from the best to the worst by Copilot. In this paper, we refer to the first suggestion as Rank 0, the second as Rank 1 an so on. In this figure, only Rank 0 and Rank 1 suggestions are shown.}
    \label{fig:copilot_solutions}
\end{figure}

As Copilot does not provide an API, the only way to invoke it is by pressing respective keys from the keyboard, namely Ctrl + Enter. We automate this action using AppleScript. Once Copilot generates suggestions, they are copied and saved to the file again with the help of AppleScript. At the end of this step, we have a file with all of the Copilot suggestions for a given problem separated with ======= lines. Figure \ref{fig:copilot_solutions} shows an example of such a file.
During our research, we noticed that Copilot sometimes does not generate any suggestions. Therefore, for each problem, we invoke Copilot $5$ times at maximum. If, by that time, Copilot does not manage to generate any suggestions, we skip the problem and do not include it in our dataset. 
Moreover, Copilot's number of generated suggestions is not fixed and can change from problem to problem. In general, the more solutions are generated by Copilot for a given problem, the more likely one of them is correct. We investigate both of these aspects in Section~\ref{sec:results} addressing \textbf{RQ1}.

\subsubsection{Form a solution and submit to LeetCode}
In this step, we iterate through all of the Copilot suggestions and insert them in the proper place of the solution template using AppleScript, followed by submitting the resulting code to LeetCode via their API.
After the submission, we get a submission ID, with the help of which we can check the solution status. Typically, it takes a couple of seconds for LeetCode to check our solution.

LeetCode limits how many submissions one can make per certain period of time. To be less affected by the rate limits, we created multiple LeetCode accounts, and while some of them were blocked, the others were used.

\subsubsection{Check the submission result and save it}
\begin{figure*}
    \centering
    \includegraphics[scale=0.49]{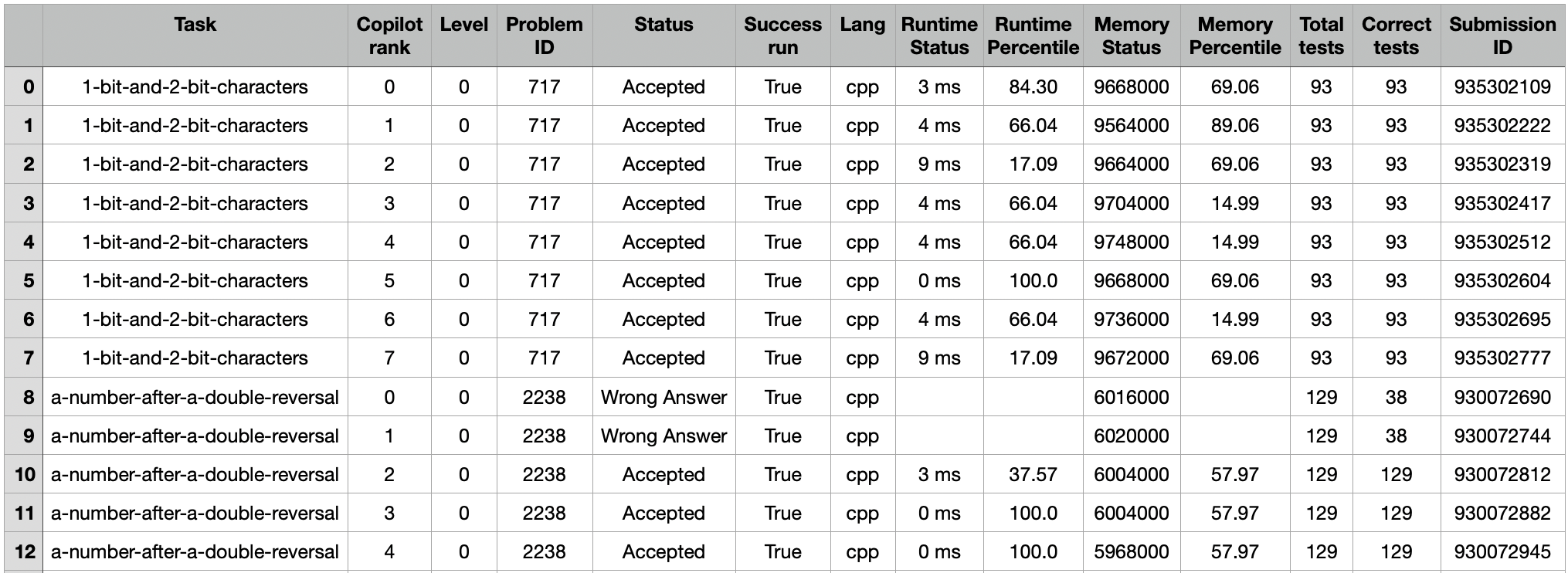}
    \caption{The head of the table with the submission results. The full table has over 50000 rows and is publicly available with the code.}
    \label{fig:grand_table_head}
\end{figure*}

Once LeetCode evaluates our submission, we fetch the detailed submission results via their API. Every submission contains information about
the submission status, the total number of tests for the problem and the number of passed tests. Moreover, when a solution is \textit{Accepted}, LeetCode provides information about how well it performed compared to all the other successful submissions on LeetCode made for the same problem. LeetCode evaluates runtime and memory consumption and provides us with the program's absolute numbers of time/memory consumption and a runtime/memory percentile to compare our submission with others.
The head of the table with the submission results is shown in Figure~\ref{fig:grand_table_head}. The full data is publicly available with the code at \url{https://github.com/IljaSir/CopilotSolverForLeetCode}.

\section{Results} \label{sec:results}
In this section, we present our results. We divide them into subsections, answering the research questions stated in the introduction.

\subsection*{RQ1: How reliable is Copilot in the code generation phase?}\label{subsec:RELIABLE}
As mentioned in Section~\ref{subsec:invoke_copilot_and_save_solutions}, sometimes Copilot fails to generate a solution on the first try. Therefore, we give $5$ attempts to Copilot to generate the solution for a given problem. If Copilot fails to do so, we skip this problem. The statistics about how many problems were skipped due to that reason are shown in Table~\ref{tab:failed}. It can be seen that for C++ and Java it rarely happens. For Rust, it largely depends on the difficulty of the problem, i.e. Copilot is more likely to fail to generate a solution for a harder problem, as one could expect. However, the language with the largest number of failed generation attempts is Python3, resulting in $469$ problems across all difficulty levels without a generated solution. 

In Section~\ref{subsec:invoke_copilot_and_save_solutions}, it was also mentioned that Copilot can generate a different number of suggestions depending on a given problem.
Table~\ref{tab:avg_k} shows the average amount of suggestions Copilot generated for a LeetCode problem depending on the difficulty level and the programming language. It can be seen that the most number of suggestions are generated for Java and C++ problems with around $9$ generated solutions regardless of the difficulty of the problem. On the other hand, Rust has a different tendency. Copilot tends to generate more suggestions for easier problems. Python3, however, performs the worst and has a stable average of around $6$ suggestions per problem regardless of the difficulty of the problem. This result is consistent with Table~\ref{tab:failed}, where Copilot also struggled to generate Python3 code.

We discuss our assumptions of why Copilot might have difficulties in generating Python3 code in Section~\ref{subsec:python_poor_perf}.

\bgroup
\def\arraystretch{1.2}
\begin{table}[]
\caption{Number of problems that were skipped as Copilot failed to generate any solutions for them.}
\begin{center}
\begin{tabular}{|c|c|c|c|c|}
\hline
\rowcolor[HTML]{B0B3B2} 
                               & \textbf{Java} & \textbf{C++} & \textbf{Python3} & \textbf{Rust} \\
\hline
\cellcolor[HTML]{D4D4D4}\textbf{Easy}   & $0$    & $0$   & $72$     & $4$    \\
\hline
\cellcolor[HTML]{D4D4D4}\textbf{Medium} & $4$    & $3$   & $210$    & $75$   \\
\hline
\cellcolor[HTML]{D4D4D4}\textbf{Hard}   & $1$    & $4$   & $187$    & $167$  \\
\hline
\rowcolor[HTML]{F2F2F2} 
\cellcolor[HTML]{D4D4D4}\textbf{Total}  & \textbf{5}    & \textbf{7}  & \textbf{469}    & \textbf{246} \\
\hline
\end{tabular}
\label{tab:failed}
\end{center}
\end{table}
\egroup

\bgroup
\def\arraystretch{1.2}
\begin{table}[]
\caption{Average number of generated suggestions per LeetCode problem.}
\begin{center}
\begin{tabular}{|c|c|c|c|c|}
\hline
\rowcolor[HTML]{B0B3B2} 
                               & \textbf{Java} & \textbf{C++} & \textbf{Python3} & \textbf{Rust} \\
\hline
\cellcolor[HTML]{D4D4D4}\textbf{Easy}   &   $8.7 \pm 2.2$  &  $9.1 \pm 1.6$  &  $5.4 \pm 3.2$    &  $8.8 \pm 2.0$   \\
\hline
\cellcolor[HTML]{D4D4D4}\textbf{Medium} &   $10.0 \pm 2.5$  &  $9.6 \pm 2.5$  &   $6.0 \pm 3.6$  &  $7.5 \pm 3.1$  \\
\hline
\cellcolor[HTML]{D4D4D4}\textbf{Hard}   &   $9.4 \pm 2.2$  &  $8.7 \pm 3.8$  &  $6.0 \pm 3.4$ &  $5.6 \pm 3.2$ \\
\hline
\end{tabular}
\label{tab:avg_k}
\end{center}
\end{table}
\egroup

\subsection*{RQ2: How correct are Copilot's solutions?}
To check the correctness of the solutions, we fully rely on LeetCode's internal tests, which are not publicly available. However, LeetCode gives information about the number of passed tests by the generated solution as well as the status of that solution. We distinguish $3$ cases: ``Correct" means that all tests were passed by the solution, ``Partially Correct" means that some but not all the tests were passed, and ``Incorrect" implies that the submitted code passed none. Here we consider Copilot's solution for the problem to be Correct if Copilot generated at least one suggestion that passes all the tests. Similarly, the solution is considered ``Partially Correct'' if it is not ``Correct'' and at least one of the Copilot's suggestions passed some tests. Table~\ref{tab:table_1} summarizes the overall correctness of Copilot's solutions for each language. It can be seen that Copilot provided correct solutions for most of the LeetCode problems for each language. Also, the number of partially correct solutions is quite high, leaving the number of incorrect solutions in a clear minority.

\subsubsection{How does the correctness of Copilot's solution depend on the programming language?}
Table~\ref{tab:table_1} shows that the acceptance rate for Copilot's solutions in C++ and Java is better than in Python3 and Rust. It is unsurprising to see C++ and Java performing so well, as those languages have been around for ages and, hence, have much more code available on GitHub that was used during Copilot's training phase. Rust, however, is a relatively modern language and does not have such a huge codebase. Hence, Copilot performs much worse for Rust than for C++ and Java. Nevertheless, the biggest surprise was the relatively poor performance of Python3, which has been one of the most popular languages on GitHub for the last $10$ years and the most popular one since $2019$~\cite{GitHut}. However, the performance of Python3 can be explained by the results from Table~\ref{tab:avg_k}. As fewer suggestions are generated for Python3, it is less likely to get the correct one. Similar results were observed by Döderlein et al.~\cite{döderlein2023piloting}, where they showed that Codex performed much better on the HumanEval dataset when it was asked to output $10$ different suggestions instead of one.

\bgroup
\def\arraystretch{1.5}
\begin{table}[]
\caption{Submission results for each language.}
\resizebox{\linewidth}{!} {
\begin{tabular}{|c|c|c|c|c|}
\hline
\rowcolor[HTML]{B0B3B2} 
\textbf{Language} & \textbf{Total Problems}  & \textbf{Correct}        & \textbf{Partially Correct} & \textbf{Incorrect}    \\
\hline
\cellcolor[HTML]{D4D4D4}\textbf{Java}     & 1735 &  1331 (75.66\%) & 375 (22.46\%)     & 29 (1.88\%)  \\
\hline
\cellcolor[HTML]{D4D4D4}\textbf{C++}      & 1751 &  1305 (73.33\%) & 400 (23.81\%)     & 46 (2.86\%)  \\
\hline
\cellcolor[HTML]{D4D4D4}\textbf{Python3}  & 1284 & 886 (66.92\%)  & 290 (24.46\%)     & 108 (8.62\%) \\
\hline
\cellcolor[HTML]{D4D4D4}\textbf{Rust}     & 1511 &  1011 (62.23\%) & 383 (28.51\%)     & 117 (9.27\%) \\
\hline
\end{tabular}
}
\label{tab:table_1}
\end{table}
\egroup

\subsubsection{How does the correctness of Copilot's solution depend on the difficulty of the problem?}
In Table~\ref{tab:table_2}, we analyse a level-by-level distribution of submissions. Overall, the number of correct solutions decreases, and the number of incorrect solutions increases when problems become harder.
This result is quite intuitive, but it proves that hard problems for humans are also hard for Copilot and vice versa: easy problems for humans are easy for Copilot.
In addition, similar tendencies to Table~\ref{tab:table_1} can be observed in Table~\ref{tab:table_2}. Namely, Copilot generates more correct solutions in C++ and Java than in Python3 and Rust for every level.

\bgroup
\def\arraystretch{1.2}
\begin{table}[]
\caption{Level by level submission results for each language.}
\resizebox{\linewidth}{!} {
\begin{tabular}{|c|c|c|c|c|c|}
\hline
\rowcolor[HTML]{B0B3B2} 
\textbf{Level}                                            & \textbf{Language} & \textbf{Total} & \textbf{Correct}       & \textbf{Partially Correct} & \textbf{Incorrect}    \\
\hline
\cellcolor[HTML]{D4D4D4}                         & \textbf{Java}     & $443$   & $406$ ($91.65$\%) & $34$ ($7.67$\%)      & $3$ ($0.68$\%)   \\
\cellcolor[HTML]{D4D4D4}                         & \textbf{C++}      & $443$   & $406$ ($91.65$\%) & $35$ ($7.9$\%)        & $2$ ($0.45$\%)   \\
\cellcolor[HTML]{D4D4D4}                         & \textbf{Python3}  & $368$   & $308$ ($83.7$\%)  & $42$ ($11.41$\%)      & $18$ ($4.89$\%)  \\
\multirow{-4}{*}{\cellcolor[HTML]{D4D4D4}\textbf{Easy}}   & \textbf{Rust}     & $438$   & $392$ ($89.5$\%)  & $38$ ($8.68$\%)       & $8$ ($1.83$\%)   \\
\hline
\cellcolor[HTML]{D4D4D4}                         & \textbf{Java}     & $872$   & $688$ ($78.9$\%)  & $174$ ($19.95$\%)     & $10$ ($1.15$\%)  \\
\cellcolor[HTML]{D4D4D4}                         & \textbf{C++}      & $881$   & $681$ ($77.3$\%)  & $182$ ($20.66$\%)     & $18$ ($2.04$\%)  \\
\cellcolor[HTML]{D4D4D4}                         & \textbf{Python3}  & $672$   & $459$ ($68.3$\%)  & $152$ ($22.62$\%)     & $61$ ($9.08$\%)  \\
\multirow{-4}{*}{\cellcolor[HTML]{D4D4D4}\textbf{Medium}} & \textbf{Rust}     & $809$   & $538$ ($66.5$\%)  & $211$ ($26.08$\%)     & $60$ ($7.42$\%)  \\
\hline
\cellcolor[HTML]{D4D4D4}                         & \textbf{Java}     & $420$   & $237$ ($56.43$\%) & $167$ ($39.76$\%)     & $16$ ($3.81$\%)  \\
\cellcolor[HTML]{D4D4D4}                         & \textbf{C++}      & $427$   & $218$ ($51.05$\%) & $183$ ($42.86$\%)     & $26$ ($6.09$\%)  \\
\cellcolor[HTML]{D4D4D4}                         & \textbf{Python3}  & $244$   & $119$ ($48.77$\%) & $96$ ($39.34$\%)      & $29$ ($11.89$\%) \\
\multirow{-4}{*}{\cellcolor[HTML]{D4D4D4}\textbf{Hard}}   & \textbf{Rust}     & $264$   & $81$ ($30.68$\%)  & $134$ ($50.76$\%)     & $49$ ($18.56$\%) \\
\hline
\end{tabular}
}
\label{tab:table_2}
\end{table}
\egroup

\subsubsection{How does the correctness of Copilot's solution depend on its rank in the code generation phase?}
Copilot provides multiple suggestions for every problem. In this subsection, we investigate how well each suggestion performs. It has to be noted that Copilot does not always generate the same amount of solutions for every problem, i.e., the solutions with rank $0$ are always present. However, the solutions with rank $i$ are only present if Copilot managed to generate at least $i+1$ solutions. Therefore, for each rank, we provide information about an absolute value of correct solutions and a percentage of correct solutions among all the generated solutions.
The results are summarized in Table~\ref{tab:table_3}. For every row, the number highlighted in green is the maximum number of correct solutions, and in orange - it is the maximum percentage of correct solutions among all the generated solutions for the current rank (we call it the relative best rank).
As one might have expected, Copilot's main suggestion (rank $0$) almost always gives the highest absolute value of the correct solutions. In some rare cases, we get more correct solutions from rank $1$. 
However, we have much more variance for the relative best rank. Rank $0$ still gives more correct answers for C++ and Java, but this is not true for Python3 and Rust. For these languages, the relative best rank varies from $0$ to $8$ and has no strong correlation between levels.

Another statistic that can be seen from Table~\ref{tab:table_3} is the difference between the overall correctness rate per all the suggestions and the correctness rate of the first suggestion, which is between 10\% and 30\% depending on the language and the difficulty of the problem.

We can conclude that experiments have shown that rank $0$ solutions are not always the best ones and it might be useful to check solutions with other ranks when using Copilot.

\bgroup
\def\arraystretch{1.5}
\begin{table*}[h!]
\caption{Success statistics of submission results for the top 10 Copilot ranks.}
\label{tab:table_3}
\resizebox{\linewidth}{!} {%
\begin{tabular}{|c|c|c|c|c|c|c|c|c|c|c|c|c|}
\hline
\rowcolor[HTML]{B0B3B2} 
\textbf{Level}                                            & \textbf{Language} & \textbf{Total}                                                        & \textbf{Rank 0}                                                      & \textbf{Rank 1}                                                      & \textbf{Rank 2}                                                      & \textbf{Rank 3}                                                      & \textbf{Rank 4}                                                      & \textbf{Rank 5}                                                      & \textbf{Rank 6}                                                      & \textbf{Rank 7}                                                      & \textbf{Rank 8}                                                      & \textbf{Rank 9}                                                      \\
\hline

\cellcolor[HTML]{D4D4D4}                         & \textbf{Java}  & \textbf{406/443 (91.6\%)}  & \textcolor{teal}{\textbf{352}}/443 (\textcolor{orange}{\textbf{79.5\%}})                                             & 348/443 (78.6\%)                                             & 350/440 (79.5\%)                                             & 340/429 (79.3\%)                                             & 325/420 (77.4\%)                                             & 296/402 (73.6\%)                                             & 260/376 (69.1\%)                                             & 223/337 (66.2\%)                                             & 186/281 (66.2\%)                                             & 119/208 (57.2\%)                                             \\
\cellcolor[HTML]{D4D4D4}                         & \textbf{C++}      & \textbf{406/443 (91.6\%)} & \textcolor{teal}{\textbf{355}}/443 (\textcolor{orange}{\textbf{80.1\%}})&347/442 (78.5\%)& 346/439  (78.8\%) & 342/435 (78.6\%) & 342/431 (79.4\%) & 319/421 (75.8\%) & 305/406 (75.1\%) & 275/381 (72.2\%) & 225/341 (66.0\%) & 173/269 (64.3\%) \\
\cellcolor[HTML]{D4D4D4}                         &      \textbf{Python3}     & \textbf{308/368 (83.7\%)}                                        & \textcolor{teal}{\textbf{257}}/368 (\textcolor{orange}{\textbf{69.8\%}})                                             & 214/350 (61.1\%)                                             & 173/270 (64.1\%)                                             & 148/230 (64.3\%)                                             & 132/202 (65.3\%)                                             & 93/169 (55.0\%)                                              & 73/142 (51.4\%)                                              & 49/111 (44.1\%)                                              & 40/83 (48.2\%)                                               & 17/45 (37.8\%)                                               \\
\multirow{-4}{*}{\cellcolor[HTML]{D4D4D4}\textbf{Easy}}   & \textbf{Rust}     & \textbf{392/438 (89.5\%)}                                             & \textcolor{teal}{\textbf{322}}/438 (73.5\%)                                             & 311/436 (71.3\%)                                             & 312/423 (\textcolor{orange}{\textbf{73.8\%}})                                             & 283/419 (67.5\%)                                             & 279/411 (67.9\%)                                             & 272/401 (67.8\%)                                             & 259/386 (67.1\%)                                             & 237/364 (65.1\%)                                             & 173/324 (53.4\%)                                             & 116/249 (46.6\%)                                             \\
\hline

\cellcolor[HTML]{D4D4D4}                         &                                      \textbf{Java}  & \textbf{688/872 (78.9\%)}        & \textcolor{teal}{\textbf{534}}/872 (\textcolor{orange}{\textbf{61.2\%}})                                             & 508/872 (58.3\%)                                             & 491/871 (56.4\%)                                             & 465/866 (53.7\%)                                             & 452/856 (52.8\%)                                             & 417/846 (49.3\%)                                             & 403/828 (48.7\%)                                             & 355/793 (44.8\%)                                             & 324/744 (43.5\%)                                             & 232/630 (36.8\%)                                             \\
\cellcolor[HTML]{D4D4D4}                         & \textbf{C++}      & \textbf{681/881 (77.3\%)}                                             & \textcolor{teal}{\textbf{521}}/881 (\textcolor{orange}{\textbf{59.1\%}})                                             & 505/875 (57.7\%)                                             & 504/867 (58.1\%)                                             & 468/856 (54.7\%)                                             & 463/850 (54.5\%)                                             & 439/832 (52.8\%)                                             & 427/808 (52.8\%)                                             & 392/781 (50.2\%)                                             & 327/724 (45.2\%)                                             & 275/584 (47.1\%)                                             \\
\cellcolor[HTML]{D4D4D4}                         &      \textbf{Python3}     & \textbf{459/672 (68.3\%)}                                        & \textcolor{teal}{\textbf{387}}/672 (57.6\%)                                             & 344/630 (54.6\%)                                             & 295/493 (59.8\%)                                             & 263/430 (61.2\%)                                             & 259/384 (67.4\%)                                             & 223/342 (65.2\%)                                             & 208/307 (\textcolor{orange}{\textbf{67.8\%}})                                             & 182/281 (64.8\%)                                             & 123/229 (53.7\%)                                             & 82/153 (53.6\%)                                              \\
\multirow{-4}{*}{\cellcolor[HTML]{D4D4D4}\textbf{Medium}} & \textbf{Rust}     & \textbf{538/809 (66.5\%)}                                             & 320/809 (39.6\%)                                             & \textcolor{teal}{\textbf{361}}/804 (44.9\%)                                             & 322/709 (45.4\%)                                             & 321/652 (\textcolor{orange}{\textbf{49.2\%}})                                             & 269/621 (43.3\%)                                             & 228/585 (39.0\%)                                             & 239/552 (43.3\%)                                             & 210/511 (41.1\%)                                             & 176/447 (39.4\%)                                             & 111/332 (33.4\%)                                             \\
\hline
\cellcolor[HTML]{D4D4D4}                         &         \textbf{Java}  & \textbf{237/420 (56.4\%)}                                    & \textcolor{teal}{\textbf{154}}/420 (36.7\%)                                             & 153/417 (\textcolor{orange}{\textbf{36.7\%}})                                             & 146/416 (35.1\%)                                             & 139/410 (33.9\%)                                             & 143/407 (35.1\%)                                             & 141/400 (35.2\%)                                             & 126/392 (32.1\%)                                             & 123/375 (32.8\%)                                             & 98/336 (29.2\%)                                              & 68/272 (25.0\%)                                              \\
\cellcolor[HTML]{D4D4D4}                         & \textbf{C++}      & \textbf{218/427 (51.0\%)}                                             & \textcolor{teal}{\textbf{139}}/427 (\textcolor{orange}{\textbf{32.6\%}})                                             & 130/407 (31.9\%)                                             & 123/389 (31.6\%)                                             & 116/375 (30.9\%)                                             & 113/368 (30.7\%)                                             & 112/355 (31.5\%)                                             & 97/334 (29.0\%)                                              & 93/311 (29.9\%)                                              & 81/269 (30.1\%)                                              & 54/193 (28.0\%)                                              \\
\cellcolor[HTML]{D4D4D4}                         &     \textbf{Python3}     & \textbf{119/244 (48.8\%)}                                          & 68/244 (27.9\%)                                              & \textcolor{teal}{\textbf{75}}/241 (31.1\%)                                              & 68/182 (37.4\%)                                              & 59/152 (38.8\%)                                              & 50/137 (36.5\%)                                              & 53/127 (\textcolor{orange}{\textbf{41.7\%}})                                              & 42/114 (36.8\%)                                              & 32/99 (32.3\%)                                               & 23/84 (27.4\%)                                               & 13/59 (22.0\%)                                               \\
\multirow{-4}{*}{\cellcolor[HTML]{D4D4D4}\textbf{Hard}}   & \textbf{Rust}     & \textbf{81/264 (30.7\%)}                                              & 40/264 (15.2\%)                                              & \textcolor{teal}{\textbf{50}}/255 (19.6\%)                                              & 45/192 (23.4\%)                                              & 38/166 (22.9\%)                                              & 31/145 (21.4\%)                                              & 32/125 (25.6\%)                                              & 26/107 (24.3\%)                                              & 29/97 (29.9\%)                                               & 23/76 (\textcolor{orange}{\textbf{30.3\%}})                                               & 8/52 (15.4\%)       \\
\hline
\end{tabular}
}
\end{table*}
\egroup

\subsubsection{How does the correctness of Copilot's solution depend on the topic of the problem?}

Each problem on LeetCode is assigned to at least one topic. The topic tells the user the category that this problem belongs to. For example, LeetCode has topics such as \textit{Array}, \textit{String}, \textit{Math} and \textit{Tree}. In total, LeetCode distinguishes $64$ different topics. Most of the time, however, the problem contains multiple topics, one broader than the others. For example, if the problem has two topics, namely \textit{Array} and \textit{Binary search}, the user can conclude that this problem requires building an algorithm based on the binary search for the array. 

The correctness of Copilot's solutions depending on the topic is presented in Table~\ref{tab:table_6}. We cut the table to have only the topics from our dataset with more than $5$ problems per language. For each language, we also add a column that shows the average difficulty of the problems for the given topic. We do that for every language because a set of problems in our dataset for a given topic might not be the same for every language. It could happen if Copilot failed to generate suggestions for one language but did generate for another.

\begin{figure}
    \centering
    \includegraphics[scale=0.5]{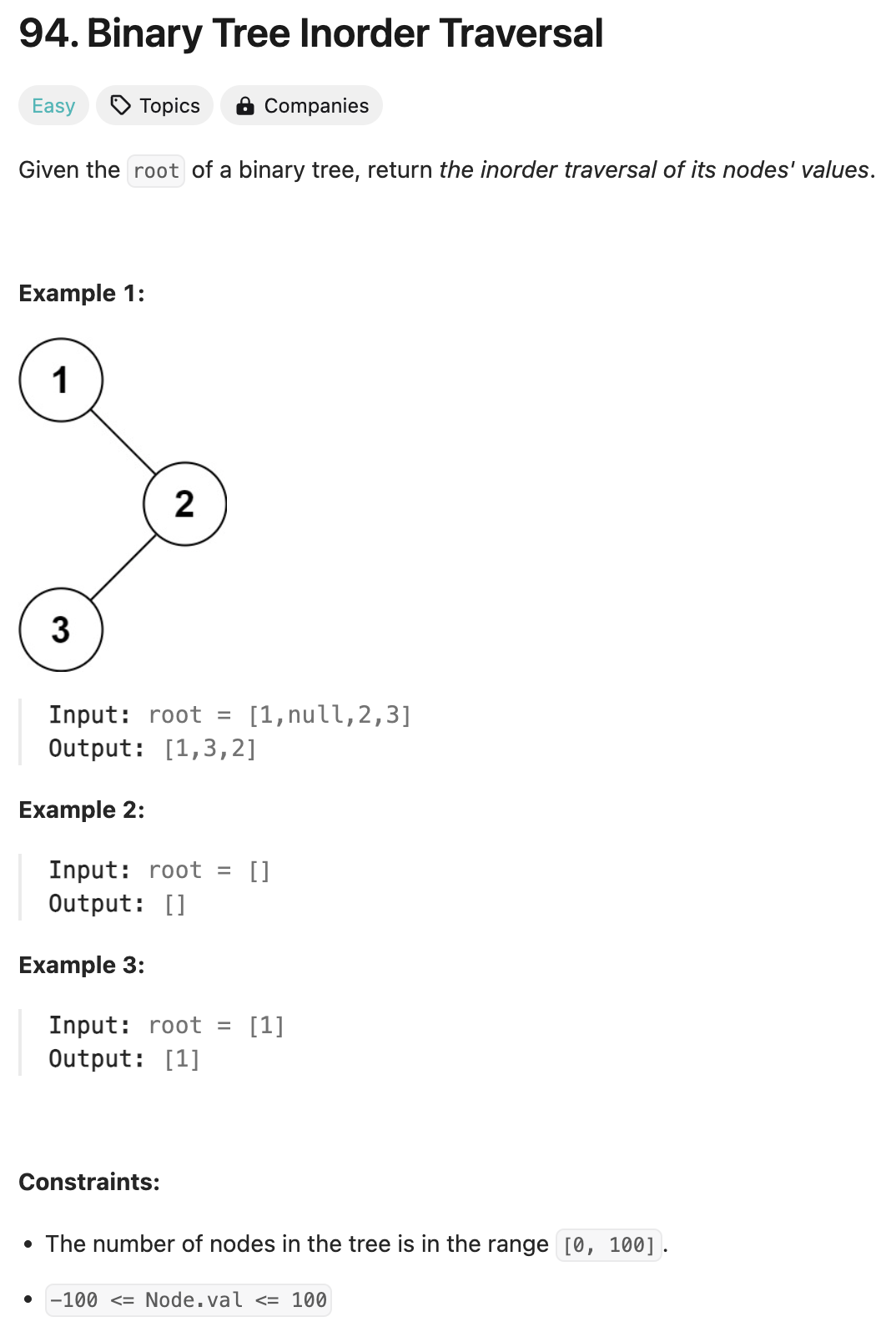}
    \caption{An example of a problem which has an image in the description. This image is aimed to help to understand the examples below. In our research Copilot had no access to information from the images. Therefore, it might be why Copilot performed worse for the problems containing images in their description.}
    \label{fig:Tree_problem_example}
\end{figure}

We can see that the most popular topics for LeetCode problems are \textit{Array}, \textit{String} and \textit{Math}. In each of them, Copilot can solve more than $70\%$ of problems on average among all programming languages. The best correctness rate is shown for \textit{Bucket Sort} problems with a correctness rate of over $95\%$. However, the most difficult topic for Copilot is \textit{Tree}, with the correctness rate of $34.2\%$. Looking deeper into \textit{Tree} problems, we noticed that most contain images as part of their description. An example of such a problem is shown in Figure~\ref{fig:Tree_problem_example}.
Those images contain important information to understand the text part of examples. As mentioned in Section~\ref{subsec:request_problem_content}, we only provided the text part of the problem description to Copilot in our research. Therefore, we suspect that the absence of information from the images made Copilot perform the worst on the problems with the Tree topic.

Table~\ref{tab:table_6} also shows that Java and C++ perform better for most topics than Python3 and Rust, which is consistent with the result we shared in the previous subsections. Nevertheless, there are a couple of exceptions. For example, for the topic \textit{Recursion}, Python3 performs better than all the other languages and the \textit{Game Theory} problems are best solved in Rust.

\bgroup
\begin{table*}[]
\caption{Correctness of Copilot's solutions depending on the topic of the given problem. Only the topics with more than 5 problems per language are shown.}
\label{tab:table_6}
\resizebox{\textwidth}{!}{
\begin{tabular}{|l|c|c|c|c|c|c|c|c|c|}
\hline
\rowcolor[HTML]{B0B3B2}
 & \textbf{Total} & \multicolumn{2}{c|}{\textbf{C++}} &  \multicolumn{2}{c|}{\textbf{Java}} &  \multicolumn{2}{c|}{\textbf{Python3}} & \multicolumn{2}{c|}{\textbf{Rust}} \\
\rowcolor[HTML]{B0B3B2}
\multirow{-2}{*}{\textbf{Topics}} & \textbf{correctness rate} & \textbf{Difficulty} & \textbf{Correctness rate} &  \textbf{Difficulty}& \textbf{Correctness rate} &  \textbf{Difficulty} & \textbf{Correctness rate} & \textbf{Difficulty} & \textbf{Correctness rate} \\
\hline
\cellcolor[HTML]{D4D4D4}\textbf{Array} & \textbf{2774/3935 (70.5\%)} & 2.0 & 807/1097 (73.6\%) & 2.0 & 819/1087 (75.3\%) & 1.9 & 541/809 (66.9\%) & 1.9 & 607/942 (64.4\%) \\
\hline
\cellcolor[HTML]{D4D4D4}\textbf{String} & \textbf{1376/1824 (75.4\%)} & 1.9 & 380/495 (76.8\%) & 1.9 & 395/493 (80.1\%) & 1.8 & 281/393 (71.5\%) & 1.8 & 320/443 (72.2\%) \\
\hline
\cellcolor[HTML]{D4D4D4}\textbf{Math} & \textbf{899/1249 (72.0\%)} & 2.0 & 262/349 (75.1\%) & 2.0 & 263/348 (75.6\%) & 1.9 & 173/251 (68.9\%) & 1.8 & 201/301 (66.8\%) \\
\hline
\cellcolor[HTML]{D4D4D4}\textbf{Hash Table} & \textbf{874/1221 (71.6\%)} & 1.8 & 253/333 (76.0\%) & 1.8 & 256/331 (77.3\%) & 1.8 & 175/263 (66.5\%) & 1.7 & 190/294 (64.6\%) \\
\hline
\cellcolor[HTML]{D4D4D4}\textbf{Dynamic Programming} & \textbf{808/1205 (67.1\%)} & 2.5 & 243/356 (68.3\%) & 2.5 & 258/350 (73.7\%) & 2.3 & 142/224 (63.4\%) & 2.3 & 165/275 (60.0\%) \\
\hline
\cellcolor[HTML]{D4D4D4}\textbf{Sorting} & \textbf{685/986 (69.5\%)} & 2.0 & 201/273 (73.6\%) & 1.9 & 198/271 (73.1\%) & 1.9 & 136/201 (67.7\%) & 1.9 & 150/241 (62.2\%) \\
\hline
\cellcolor[HTML]{D4D4D4}\textbf{Greedy} & \textbf{552/908 (60.8\%)} & 2.1 & 163/257 (63.4\%) & 2.1 & 170/251 (67.7\%) & 2.0 & 97/172 (56.4\%) & 2.0 & 122/228 (53.5\%) \\
\hline
\cellcolor[HTML]{D4D4D4}\textbf{Binary Search} & \textbf{398/612 (65.0\%)} & 2.2 & 115/169 (68.0\%) & 2.2 & 113/166 (68.1\%) & 2.1 & 85/129 (65.9\%) & 2.2 & 85/148 (57.4\%) \\
\hline
\cellcolor[HTML]{D4D4D4}\textbf{Matrix} & \textbf{394/529 (74.5\%)} & 2.1 & 120/156 (76.9\%) & 2.1 & 124/155 (80.0\%) & 2.1 & 66/99 (66.7\%) & 2.0 & 84/119 (70.6\%) \\
\hline
\cellcolor[HTML]{D4D4D4}\textbf{Bit Manipulation} & \textbf{339/493 (68.8\%)} & 2.1 & 102/142 (71.8\%) & 2.1 & 102/137 (74.5\%) & 2.0 & 63/101 (62.4\%) & 1.9 & 72/113 (63.7\%) \\
\hline
\cellcolor[HTML]{D4D4D4}\textbf{Two Pointers} & \textbf{366/475 (77.1\%)} & 1.8 & 100/126 (79.4\%) & 1.8 & 100/125 (80.0\%) & 1.7 & 84/106 (79.2\%) & 1.7 & 82/118 (69.5\%) \\
\hline
\cellcolor[HTML]{D4D4D4}\textbf{Breadth-First Search} & \textbf{264/384 (68.8\%)} & 2.4 & 84/117 (71.8\%) & 2.3 & 86/116 (74.1\%) & 2.3 & 49/71 (69.0\%) & 2.2 & 45/80 (56.2\%) \\
\hline
\cellcolor[HTML]{D4D4D4}\textbf{Simulation} & \textbf{287/378 (75.9\%)} & 1.6 & 83/102 (81.4\%) & 1.6 & 80/102 (78.4\%) & 1.6 & 52/78 (66.7\%) & 1.6 & 72/96 (75.0\%) \\
\hline
\cellcolor[HTML]{D4D4D4}\textbf{Prefix Sum} & \textbf{205/372 (55.1\%)} & 2.2 & 63/106 (59.4\%) & 2.2 & 62/103 (60.2\%) & 2.1 & 33/74 (44.6\%) & 2.1 & 47/89 (52.8\%) \\
\hline
\cellcolor[HTML]{D4D4D4}\textbf{Heap (Priority Queue)} & \textbf{224/366 (61.2\%)} & 2.3 & 73/108 (67.6\%) & 2.2 & 68/104 (65.4\%) & 2.2 & 41/69 (59.4\%) & 2.2 & 42/85 (49.4\%) \\
\hline
\cellcolor[HTML]{D4D4D4}\textbf{Stack} & \textbf{242/325 (74.5\%)} & 2.2 & 67/88 (76.1\%) & 2.2 & 72/88 (81.8\%) & 2.1 & 48/69 (69.6\%) & 2.1 & 55/80 (68.8\%) \\
\hline
\cellcolor[HTML]{D4D4D4}\textbf{Depth-First Search} & \textbf{228/323 (70.6\%)} & 2.3 & 70/97 (72.2\%) & 2.3 & 72/97 (74.2\%) & 2.2 & 45/57 (78.9\%) & 2.2 & 41/72 (56.9\%) \\
\hline
\cellcolor[HTML]{D4D4D4}\textbf{Counting} & \textbf{228/316 (72.2\%)} & 1.7 & 65/85 (76.5\%) & 1.7 & 63/84 (75.0\%) & 1.6 & 46/70 (65.7\%) & 1.6 & 54/77 (70.1\%) \\
\hline
\cellcolor[HTML]{D4D4D4}\textbf{Graph} & \textbf{170/295 (57.6\%)} & 2.5 & 57/94 (60.6\%) & 2.5 & 58/94 (61.7\%) & 2.4 & 27/49 (55.1\%) & 2.3 & 28/58 (48.3\%) \\
\hline
\cellcolor[HTML]{D4D4D4}\textbf{Backtracking} & \textbf{188/266 (70.7\%)} & 2.3 & 56/75 (74.7\%) & 2.3 & 58/75 (77.3\%) & 2.2 & 40/57 (70.2\%) & 2.2 & 34/59 (57.6\%) \\
\hline
\cellcolor[HTML]{D4D4D4}\textbf{Sliding Window} & \textbf{182/265 (68.7\%)} & 2.2 & 55/72 (76.4\%) & 2.2 & 53/71 (74.6\%) & 2.1 & 35/57 (61.4\%) & 2.1 & 39/65 (60.0\%) \\
\hline
\cellcolor[HTML]{D4D4D4}\textbf{Union Find} & \textbf{115/182 (63.2\%)} & 2.5 & 36/56 (64.3\%) & 2.5 & 41/56 (73.2\%) & 2.4 & 21/32 (65.6\%) & 2.3 & 17/38 (44.7\%) \\
\hline
\cellcolor[HTML]{D4D4D4}\textbf{Enumeration} & \textbf{68/146 (46.6\%)} & 2.0 & 21/42 (50.0\%) & 2.0 & 20/42 (47.6\%) & 2.0 & 13/30 (43.3\%) & 1.8 & 14/32 (43.8\%) \\
\hline
\cellcolor[HTML]{D4D4D4}\textbf{Monotonic Stack} & \textbf{104/133 (78.2\%)} & 2.3 & 30/36 (83.3\%) & 2.3 & 31/36 (86.1\%) & 2.3 & 21/27 (77.8\%) & 2.3 & 22/34 (64.7\%) \\
\hline
\cellcolor[HTML]{D4D4D4}\textbf{Number Theory} & \textbf{53/114 (46.5\%}) & 2.2 & 18/33 (54.5\%) & 2.2 & 16/33 (48.5\%) & 2.2 & 8/20 (40.0\%) & 2.1 & 11/28 (39.3\%) \\
\hline
\cellcolor[HTML]{D4D4D4}\textbf{Divide and Conquer} & \textbf{78/108 (72.2\%)} & 2.2 & 22/29 (75.9\%) & 2.2 & 23/29 (79.3\%) & 2.2 & 19/24 (79.2\%) & 2.2 & 14/26 (53.8\%) \\
\hline
\cellcolor[HTML]{D4D4D4}\textbf{Memoization} & \textbf{58/99 (58.6\%)} & 2.5 & 17/30 (56.7\%) & 2.5 & 18/30 (60.0\%) & 2.5 & 10/17 (58.8\%) & 2.4 & 13/22 (59.1\%) \\
\hline
\cellcolor[HTML]{D4D4D4}\textbf{Bitmask} & \textbf{46/96 (47.9\%)} & 2.7 & 17/33 (51.5\%) & 2.7 & 17/32 (53.1\%) & 2.5 & 7/17 (41.2\%) & 2.4 & 5/14 (35.7\%) \\
\hline
\cellcolor[HTML]{D4D4D4}\textbf{Trie} & \textbf{72/94 (76.6\%)} & 2.4 & 17/25 (68.0\%) & 2.4 & 21/25 (84.0\%) & 2.3 & 18/21 (85.7\%) & 2.4 & 16/23 (69.6\%) \\
\hline
\cellcolor[HTML]{D4D4D4}\textbf{Recursion} & \textbf{79/94 (84.0\%)} & 2.2 & 22/25 (88.0\%) & 2.3 & 21/26 (80.8\%) & 2.4 & 18/20 (90.0\%) & 2.2 & 18/23 (78.3\%) \\
\hline
\cellcolor[HTML]{D4D4D4}\textbf{Geometry} & \textbf{59/91 (64.8\%)} & 1.9 & 17/26 (65.4\%) & 1.9 & 19/26 (73.1\%) & 1.8 & 10/16 (62.5\%) & 1.8 & 13/23 (56.5\%) \\
\hline
\cellcolor[HTML]{D4D4D4}\textbf{Ordered Set} & \textbf{46/87 (52.9\%)} & 2.7 & 16/24 (66.7\%) & 2.7 & 16/24 (66.7\%) & 2.7 & 9/19 (47.4\%) & 2.8 & 5/20 (25.0\%) \\
\hline
\cellcolor[HTML]{D4D4D4}\textbf{Tree} & \textbf{27/79 (34.2\%)} & 2.6 & 10/27 (37.0\%) & 2.6 & 9/27 (33.3\%) & 2.3 & 4/9 (44.4\%) & 2.5 & 4/16 (25.0\%) \\
\hline
\cellcolor[HTML]{D4D4D4}\textbf{Topological Sort} & \textbf{39/77 (50.6\%)} & 2.6 & 13/24 (54.2\%) & 2.6 & 12/24 (50.0\%) & 2.5 & 7/15 (46.7\%) & 2.4 & 7/14 (50.0\%) \\
\hline
\cellcolor[HTML]{D4D4D4}\textbf{Queue} & \textbf{43/75 (57.3\%)} & 2.4 & 14/22 (63.6\%) & 2.4 & 13/21 (61.9\%) & 2.4 & 7/15 (46.7\%) & 2.2 & 9/17 (52.9\%) \\
\hline
\cellcolor[HTML]{D4D4D4}\textbf{Game Theory} & \textbf{58/69 (84.1\%)} & 2.2 & 16/20 (80.0\%) & 2.2 & 17/20 (85.0\%) & 2.1 & 10/12 (83.3\%) & 2.2 & 15/17 (88.2\%) \\
\hline
\cellcolor[HTML]{D4D4D4}\textbf{Binary Indexed Tree} & \textbf{32/64 (50.0\%)} & 2.7 & 9/17 (52.9\%) & 2.7 & 10/17 (58.8\%) & 2.6 & 8/14 (57.1\%) & 2.7 & 5/16 (31.2\%) \\
\hline
\cellcolor[HTML]{D4D4D4}\textbf{Segment Tree} & \textbf{28/60 (46.7\%)} & 2.9 & 8/17 (47.1\%) & 2.9 & 10/17 (58.8\%) & 2.8 & 7/12 (58.3\%) & 2.9 & 3/14 (21.4\%) \\
\hline
\cellcolor[HTML]{D4D4D4}\textbf{Combinatorics} & \textbf{29/59 (49.2\%)} & 2.6 & 10/18 (55.6\%) & 2.6 & 9/17 (52.9\%) & 2.4 & 5/9 (55.6\%) & 2.5 & 5/15 (33.3\%) \\
\hline
\cellcolor[HTML]{D4D4D4}\textbf{String Matching} & \textbf{41/53 (77.4\%)} & 2.0 & 11/15 (73.3\%) & 2.0 & 11/15 (73.3\%) & 1.7 & 9/10 (90.0\%) & 1.9 & 10/13 (76.9\%) \\
\hline
\cellcolor[HTML]{D4D4D4}\textbf{Hash Function} & \textbf{21/48 (43.8\%)} & 2.7 & 5/15 (33.3\%) & 2.7 & 7/15 (46.7\%) & 2.5 & 4/8 (50.0\%) & 2.6 & 5/10 (50.0\%) \\
\hline
\cellcolor[HTML]{D4D4D4}\textbf{Rolling Hash} & \textbf{21/44 (47.7\%)} & 2.7 & 5/14 (35.7\%) & 2.7 & 7/14 (50.0\%) & 2.4 & 4/7 (57.1\%) & 2.6 & 5/9 (55.6\%) \\
\hline
\cellcolor[HTML]{D4D4D4}\textbf{Brainteaser} & \textbf{33/41 (80.5\%)} & 1.9 & 9/11 (81.8\%) & 1.9 & 7/9 (77.8\%) & 1.9 & 8/10 (80.0\%) & 1.9 & 9/11 (81.8\%) \\
\hline
\cellcolor[HTML]{D4D4D4}\textbf{Shortest Path} & \textbf{27/41 (65.9\%)} & 2.5 & 10/13 (76.9\%) & 2.5 & 10/13 (76.9\%) & 2.3 & 2/6 (33.3\%) & 2.3 & 5/9 (55.6\%) \\
\hline
\cellcolor[HTML]{D4D4D4}\textbf{Monotonic Queue} & \textbf{20/35 (57.1\%)} & 2.7 & 7/11 (63.6\%) & 2.7 & 6/10 (60.0\%) & 2.8 & 3/6 (50.0\%) & 2.6 & 4/8 (50.0\%) \\
\hline
\cellcolor[HTML]{D4D4D4}\textbf{Merge Sort} & \textbf{14/27 (51.9\%)} & 2.9 & 4/7 (57.1\%) & 2.9 & 5/7 (71.4\%) & 2.8 & 4/6 (66.7\%) & 2.9 & 1/7 (14.3\%) \\
\hline
\cellcolor[HTML]{D4D4D4}\textbf{Quickselect} & \textbf{23/26 (88.5\%)} & 2.0 & 6/7 (85.7\%) & 2.0 & 6/7 (85.7\%) & 2.0 & 6/6 (100.0\%) & 2.0 & 5/6 (83.3\%) \\
\hline
\cellcolor[HTML]{D4D4D4}\textbf{Bucket Sort} & \textbf{23/24 (95.8\%)} & 2.3 & 6/6 (100.0\%) & 2.3 & 6/6 (100.0\%) & 2.3 & 5/6 (83.3\%) & 2.3 & 6/6 (100.0\%) \\
\hline
\end{tabular}
}
\end{table*}
\egroup

\subsection*{RQ3: How does the generated solution's time and memory efficiency compare to an average human submission?}
Finally, we would like to analyze the efficiency of Copilot's generated code. For this analysis, we rely on LeetCode's ability to compare the accepted solution's runtime and memory consumption with all the other accepted submissions for the same problem. As in most of the algorithms, there is a trade-off between optimizing runtime vs memory consumption; we decided to analyze it in two ways as follows:
\begin{itemize}
    \item Take the runtime and memory percentile mean among all the Copilot accepted solutions (all ranks) and all the problems
    \item Take the maximum values (among all of the ranks) for every problem and then take a mean of that
\end{itemize}

The results are summarized in Table \ref{tab:table_4}.
Overall, the mean of maximum runtime percentiles for each language is quite high: $77$ for Python3 and over $86$ for other languages. The same can be said about the mean of maximum memory percentiles: over $75$ for every language.
The overall means of all the accepted problems are approximately $25\%$ smaller than the maximum, but all the values are still above the $50$th percentile.
We can conclude that, in general, Copilot generates more efficient code than the average human.

\bgroup
\def\arraystretch{1.5}
\begin{table}[]
\caption{Efficiency of Copilot generated code.}
\label{tab:table_4}
\resizebox{\linewidth}{!} {
\begin{tabular}{|c|c|c|c|c|}
\hline
\rowcolor[HTML]{B0B3B2} 
\textbf{Language}                        & \begin{tabular}[c]{@{}c@{}}\textbf{Max Runtime} \\  \textbf{Percentile}\\  \textbf{(mean)}\end{tabular} & \begin{tabular}[c]{@{}c@{}}\textbf{Mean Runtime}\\  \textbf{Percentile}\\  \textbf{(mean)}\end{tabular} & \begin{tabular}[c]{@{}c@{}}\textbf{Max Memory}\\  \textbf{Percentile}\\  \textbf{(mean)}\end{tabular} & \begin{tabular}[c]{@{}c@{}}\textbf{Mean Memory}\\  \textbf{Percentile}\\  \textbf{(mean)}\end{tabular} \\
\hline
\cellcolor[HTML]{D4D4D4}\textbf{Java}    & $86.25$                                                                        & $73.46$                                                                        & $78.91$                                                                      & $53.14$                                                                       \\
\hline
\cellcolor[HTML]{D4D4D4}\textbf{C++}     & $86.16$                                                                        & $62.98$                                                                        & $80.91$                                                                      & $57.83$                                                                       \\
\hline
\cellcolor[HTML]{D4D4D4}\textbf{Python3} & $77.11$                                                                        & $59.61$                                                                        & $76.18$                                                                      & $57.43$                                                                       \\
\hline
\cellcolor[HTML]{D4D4D4}\textbf{Rust}    & $87.74$                                                                        & $73.21$                                                                        & $81.49$                                                                      & $61.70$                                                                      \\
\hline
\end{tabular}
}
\end{table}
\egroup

\section{Discussion}
\subsection{Comparison with previous works}
\label{subsec:comparison_with_prev_works}
As mentioned in Section~\ref{sec:RelatedWork}, a couple of papers did similar research on a smaller amount of problems.
Table~\ref{tab:comparison_with_other_papers} compares the results of 2 previous papers to those from this work. It should be noted that the previous works only analysed the Rank $0$ suggestions of Copilot. For the sake of fair comparison, in the column corresponding to this work, we also only consider Rank $0$ solutions. It can be seen that Copilot's results on a larger dataset lead to better performance on LeetCode problems.

\bgroup
\def\arraystretch{1.2}
\begin{table*}[]
\caption{Comparison with previous works.}
\centering
\begin{tabular}{|c|c|c|c|c|c|c|}
\hline
\rowcolor[HTML]{B0B3B2} 
    & \multicolumn{2}{c|}{\textbf{Döderlein et al.~\cite{döderlein2023piloting}}} & \multicolumn{2}{c|}{\textbf{Nguyen et al.~\cite{Copilot_Eval}}} & \multicolumn{2}{c|}{\textbf{Our work (Only Rank 0)}} \\
\hline
\rowcolor[HTML]{B0B3B2} 
 \textbf{Language} & \textbf{Total} & \textbf{Correct} & \textbf{Total} & \textbf{Correct}       & \textbf{Total} & \textbf{Correct}          \\
\hline
\cellcolor[HTML]{D4D4D4}                          \textbf{Python3}     & $300$ &  $33.7$\%  & $33$ & $42.4$\% & $1284$ & $55.4$\%       \\
\cellcolor[HTML]{D4D4D4}                          \textbf{Java}      & $300$   & $46.3$\% & $33$ &  $57.6$\% & $1735$ & $59.9$\%        \\
\cellcolor[HTML]{D4D4D4}                          \textbf{C++}  & $300$   &  $45.3$\%  & - & -  &  $1751$ & $58.0$\%  \\
\cellcolor[HTML]{D4D4D4}  \textbf{Rust}     & -   & -  & - & - &  $1511$ & $45.1$\%     \\
\hline
\end{tabular}
\label{tab:comparison_with_other_papers}
\end{table*}
\egroup

\subsection{Why does Copilot have relatively poor performance in Python3?} \label{subsec:python_poor_perf}
In Section~\ref{subsec:RELIABLE} we showed that Copilot preforms worse for Python3 than for Java and C++. It happens in both generation phase, where Copilot fails to generate any suggestions for many Python3 problems and generates less suggestions on average per problem, and in the code's correctness, where generated in Python3 code is less likely to be a correct solution to a given problem than Java or C++ code. Before sharing our assumptions of why this might be the case, we note that Copilot was trained on the public GitHub repositories, which features a lot of Python3 code, as Python3 has been one of the most popular languages on GitHub for the last decade~\cite{GitHut}.

First of all, we argue that Python3 is a much easier language for the average person to start coding in than C++ and Java. This means that there might be more low-quality Python3 code written by beginners on GitHub, which might have a negative impact on the quality of code that Copilot learns to produce. Secondly, Python3 is a much less strict programming language than Java and C++. For example, it does not require programmer to set explicit types to variables. It also does not require the main function to be present in the code for it to be run successfully. Thanks to those language features, a programmer can accomplish the given task in a number of ways while achieving the same result. However, at the same time, the flexibility of Python3 makes it harder to learn by just looking at examples, which is what Copilot does during the training phase.
In conclusion, we believe that to improve results for Python3, Copilot needs to be trained more on the Python3 code, preferably of high quality. Further research is needed to verify the correctness of our assumptions.

\subsection{Threats to validity}
\subsubsection*{Correctness of Copilot's solutions} In Section~\ref{sec:results} answering to \textbf{RQ2}, we assessed the correctness of Copilot's suggestions. There, we fully rely on LeetCode tests to prove or disprove that. Those tests, in general, are not guaranteed to be comprehensive. However, LeetCode is one of the most widely used problem-solving websites, with many people using it, which is constantly evolving. Therefore, we assume that it is highly unlikely that LeetCode will misclassify an incorrect solution as a correct one.
\subsubsection*{Copilot vs Average human performance} In Section~\ref{sec:results} answering to \textbf{RQ3}, we stated that Copilot solutions perform better than an average human. Not all of the submissions to LeetCode were written by humans, as there were similar research works to the current one made in the past. However, Copilot performs significantly better than the average submission to LeetCode. Therefore, our claims might in theory not be valid in the hypothetical case where many more submissions made to LeetCode are written by other AI tools that perform worse than Copilot. However, in practice we do not expect this number of solutions to be high enough to drop the percentages from Table~\ref{tab:table_4} down to less than $50$\%.
\subsubsection*{Time\&Memory efficiency of solutions} In the same section, we measure the time and memory performance of the Copilot suggestions relying on the LeetCode internal runs. We can assume that LeetCode tests run on cloud-based architectures. Therefore, time and memory metrics might not be fully trustworthy. Separate research is needed to determine the precision of those metrics.

\subsubsection*{Recitation}
\begin{figure*}
    \centering
    \includegraphics[scale=0.45]{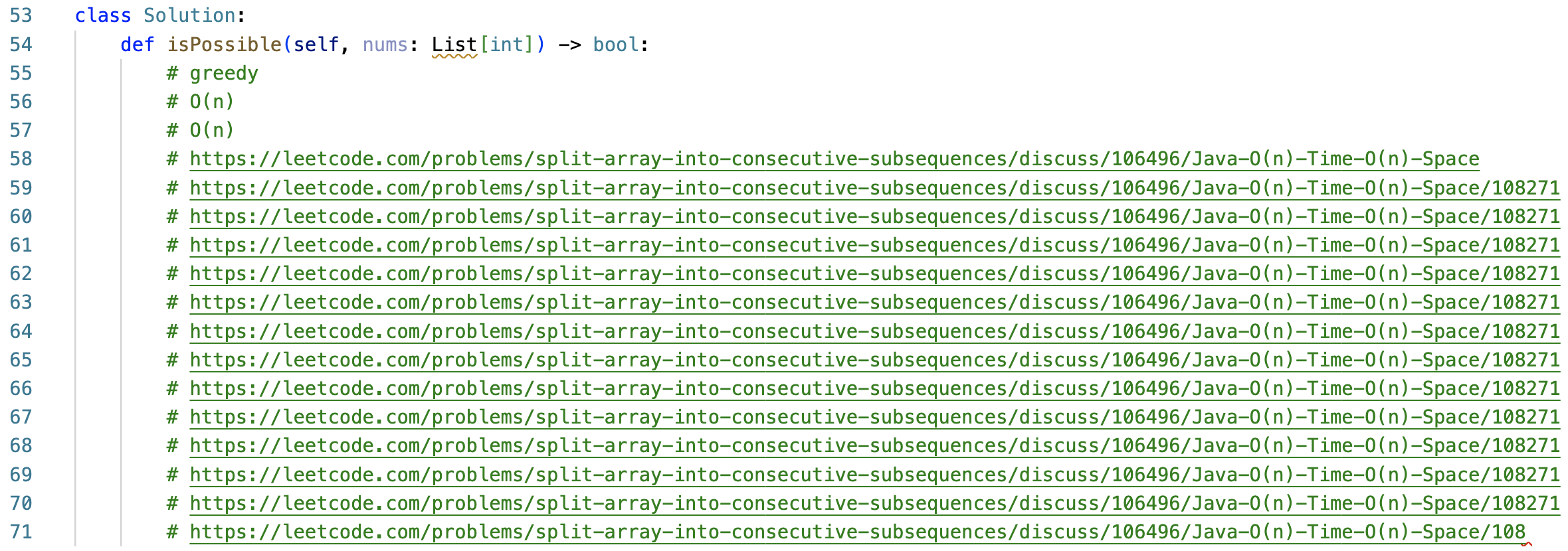}
    \caption{An example of a Copilot answer with links to LeetCode.}
    \label{fig:Strange_Copilot_response}
\end{figure*}

Recitation is one of the biggest concerns of large language models. It happens if a language model generates a chunk of training data as its output. Recitation can question the ability of those models to generalize the information instead of just repeating what they saw during the training phase. In our work, it can lead to a case where Copilot repeats some solutions for a given LeetCode problem that it saw on the web.
Initial study by GitHub~\cite{Copilot_recitation} reported around $5\%$ of recitation in Copilot outputs. A more recent study by Ciniselli et al.~\cite{ciniselli2022extent} reported a higher number of recitations, namely $10\%$. Döderlein et al.
~\cite{döderlein2023piloting} assumed that having links to LeetCode in a problem description fed to Copilot might lead to a higher success rate. In our experiments, we noticed that only $8$ problem descriptions from $1760$ had a link to LeetCode included. Despite that, we discovered that for some problems (that did not contain a link in their description), Copilot generated an output containing some (non-working) links to LeetCode. Figure~\ref{fig:Strange_Copilot_response} shows an example of such an output. This proves that for those problems, Copilot was definitely aware that they came from LeetCode. Therefore, a recitation might have happened during the suggestions generation process. In total, we saw similar outputs from Copilot for $5$ problems.
Interestingly enough, this behaviour happened only when generating solutions in Python3. A further study is needed to estimate the amount of recitation in Copilot suggestions for LeetCode problems.

\subsubsection*{Reproducibility of the results} Even though we shared the code and the data of our experiments, our results might not be fully reproducible as GitHub Copilot is a non-deterministic language model, which might lead to having different outputs in different executions on the same input.

\section{Conclusion and Future work}
In this paper, we automated the evaluation process of the code generated by GitHub Copilot. We analysed the reliability of the generation process, the code's correctness and its dependency on the programming language, Copilot's rank and the topic of the problem. We focused our research on $4$ programming languages: Java, C++, Python3 and Rust. We used LeetCode's free problem database as the evaluation dataset. In total, we asked Copilot to generate code for $1760$ problems, which resulted in over $50000$ submissions to LeetCode spread over a period of $2$ month.

We showed, that in the code generation phase, Copilot is more reliable for Java and C++, resulting in less generation failures and more suggestions per problem than Python3 and Rust. Similar results were achieved in the evaluation of the correctness: Copilot solved more LeetCode problems in Java and C++ than in Python3 and Rust, resulting in more than $72\%$ success rate across all the languages.
Moreover, we found that Copilot's suggestions' ranking system does not necessarily correlate with the best solutions' ranking, i.e., the top Copilot proposal is not always the best one. So, if Copilot generates multiple solutions, it is worth exploring them all.
In addition, we found that from all of the problem topics, \textit{Bucket Sort} has the highest correctness rate while \textit{Tree} has the lowest.

We also investigated time and memory efficiency of the generated code. We concluded that the code generated by Copilot is more efficient, both time- and memory-wise, than the code written by the average human.

In future work, it might be interesting to enhance the number of programming languages for code generation and use other datasets to check if our results can be generalized.
Our work can also be used as a point of reference for similar studies that might be conducted in the future using a newer version of Copilot to evaluate its further progress in performance.

\bibliographystyle{ieeetr}
\bibliography{references}

\begin{thebibliography}{10}

\bibitem{Copilot}
GitHub, ``{GitHub Copilot - Your AI pair programmer}.'' \url{https://github.com/features/copilot}, 2022.
\newblock [Online; accessed 6-May-2023].

\bibitem{LeetCode}
LeetCode, ``{LeetCode - The World's Leading Online Programming Learning Platform}.'' \url{https://leetcode.com/}, 2023.
\newblock [Online; accessed 6-May-2023].

\bibitem{LeetCode_questions_difficulty}
LeetCode, ``{[Not an interview question] How does leetcode assign a difficulty level to the questions ?}.'' \url{https://leetcode.com/discuss/general-discussion/136693/not-an-interview-question-how-does-leetcode-assign-a-difficulty-level-to-the-questions/}, 2016.
\newblock [Online; accessed 6-February-2024].

\bibitem{pearce2021asleep}
H.~Pearce, B.~Ahmad, B.~Tan, B.~Dolan-Gavitt, and R.~Karri, ``Asleep at the keyboard? assessing the security of github copilot's code contributions,'' 2021.

\bibitem{CWE}
MITRE, ``{CWE - Common Weakness Enumeration}.'' \url{https://cwe.mitre.org/}, 2023.
\newblock [Online; accessed 20-November-2023].

\bibitem{10.1145/3510454.3522684}
S.~Imai, ``Is github copilot a substitute for human pair-programming? an empirical study,'' in {\em Proceedings of the ACM/IEEE 44th International Conference on Software Engineering: Companion Proceedings}, ICSE '22, (New York, NY, USA), p.~319–321, Association for Computing Machinery, 2022.

\bibitem{dakhel2023github}
A.~M. Dakhel, V.~Majdinasab, A.~Nikanjam, F.~Khomh, M.~C. Desmarais, Z.~Ming, and Jiang, ``Github copilot ai pair programmer: Asset or liability?,'' 2023.

\bibitem{10.1145/3558489.3559072}
B.~Yetistiren, I.~Ozsoy, and E.~Tuzun, ``Assessing the quality of github copilot’s code generation,'' in {\em Proceedings of the 18th International Conference on Predictive Models and Data Analytics in Software Engineering}, PROMISE 2022, (New York, NY, USA), p.~62–71, Association for Computing Machinery, 2022.

\bibitem{Copilot_Eval}
N.~Nguyen and S.~Nadi, ``An empirical evaluation of github copilot's code suggestions,'' in {\em 2022 IEEE/ACM 19th International Conference on Mining Software Repositories (MSR)}, pp.~1--5, 2022.

\bibitem{humaneval}
M.~Chen, J.~Tworek, H.~Jun, Q.~Yuan, H.~P. de~Oliveira~Pinto, J.~Kaplan, H.~Edwards, Y.~Burda, N.~Joseph, G.~Brockman, A.~Ray, R.~Puri, G.~Krueger, M.~Petrov, H.~Khlaaf, G.~Sastry, P.~Mishkin, B.~Chan, S.~Gray, N.~Ryder, M.~Pavlov, A.~Power, L.~Kaiser, M.~Bavarian, C.~Winter, P.~Tillet, F.~P. Such, D.~Cummings, M.~Plappert, F.~Chantzis, E.~Barnes, A.~Herbert-Voss, W.~H. Guss, A.~Nichol, A.~Paino, N.~Tezak, J.~Tang, I.~Babuschkin, S.~Balaji, S.~Jain, W.~Saunders, C.~Hesse, A.~N. Carr, J.~Leike, J.~Achiam, V.~Misra, E.~Morikawa, A.~Radford, M.~Knight, M.~Brundage, M.~Murati, K.~Mayer, P.~Welinder, B.~McGrew, D.~Amodei, S.~McCandlish, I.~Sutskever, and W.~Zaremba, ``Evaluating large language models trained on code,'' 2021.

\bibitem{döderlein2023piloting}
J.-B. Döderlein, M.~Acher, D.~E. Khelladi, and B.~Combemale, ``Piloting copilot and codex: Hot temperature, cold prompts, or black magic?,'' 2023.

\bibitem{VS_CODE}
Microsoft, ``{Visual Studio Code - Code Editing. Redefined.}.'' \url{https://code.visualstudio.com/}, 2023.
\newblock [Online; accessed 20-November-2023].

\bibitem{python-LeetCode}
P.~Safronov, ``{Leetcode API implementation}.'' \url{https://pypi.org/project/python-leetcode/}, 2021.
\newblock [Online; accessed 6-May-2023].

\bibitem{AppleScript}
Apple, ``{AppleScript Language Guide}.'' \url{https://developer.apple.com/library/archive/documentation/AppleScript/Conceptual/AppleScriptLangGuide/introduction/ASLR_intro.html}, 2016.
\newblock [Online; accessed 11-April-2023].

\bibitem{LeetCode_API_ALL}
LeetCode, ``{List of all LeetCode problem names}.'' \url{https://leetcode.com/api/problems/all/}, 2023.
\newblock [Online; accessed 13-July-2023].

\bibitem{GitHut}
C.~Zapponi, ``{Github Language Stats}.'' \url{https://madnight.github.io/githut}.
\newblock [Online; accessed 13-July-2023].

\bibitem{Copilot_recitation}
A.~Ziegler, ``{GitHub Copilot research recitation}.'' \url{https://github.blog/2021-06-30-github-copilot-research-recitation}, 2021.
\newblock [Online; accessed 13-February-2024].

\bibitem{ciniselli2022extent}
M.~Ciniselli, L.~Pascarella, and G.~Bavota, ``To what extent do deep learning-based code recommenders generate predictions by cloning code from the training set?,'' 2022.

\end{thebibliography}

\end{document}